\def\sqr#1#2{{\vcenter{\vbox{\hrule height.#2pt\hbox{\vrule
width.#2pt height#1pt \kern#1pt\vrule width.#2pt}\hrule height.#2pt}}}}
\def\square{\mathchoice\sqr54\sqr54\sqr{2.1}3\sqr{1.5}3} 

\magnification\magstep1
\font\cst=cmr10 scaled \magstep3
\font\csc=cmr10 scaled \magstep2
\vglue 1.5cm
\centerline{\cst  On linearized gravity in the Randall-Sundrum scenario}
\vskip 1.5 true cm
\centerline{Nathalie Deruelle$^{1,2,3}$, Tom\'a\v s Dole\v zel$^{1,4}$}
\vskip 1 true cm
\centerline{$^1$\it D\'epartement d'Astrophysique Relativiste et de
Cosmologie, }
\centerline{\it UMR 8629 du Centre National de la Recherche Scientifique,}
\centerline{\it Observatoire de Paris, 92195 Meudon, France}
\medskip
\centerline{$^2$ \it Institut des Hautes Etudes Scientifiques,}
\centerline{\it 91440, Bures-sur-Yvette, France}
\medskip
\centerline{$^3$ \it Centre for  Mathematical Sciences, DAMTP, University of Cambridge,}
\centerline{\it Wilberforce Road, Cambridge, CB3 0WA, England}
\medskip
\centerline{$^4$ \it Institute of Theoretical Physics, Charles University,}
\centerline{\it V Hole\v sovi\v ck\'ach 2, 18000 Prague 8,  Czech Republic}
\medskip
\vskip 0.5cm
\centerline{May 27th 2001}

\vskip 0.5cm
{\bf Pacs Numbers}~: 98.80.Cq, 98.70.Vc

\vskip 1cm
\noindent
{\bf Abstract}

In the literature about the Randall-Sundrum scenario one finds on one hand that there exist (small) corrections
to Newton's law of gravity on the brane, and on another that the exact (and henceforth linearized) Einstein equations can
be recovered on the brane. The explanation for these seemingly contradictory results is that the behaviour of the bulk far
from the brane is different in both models. We show that explicitely in this paper.
\vfill\eject

\noindent
{\csc I. Introduction}
\medskip
There has been recently an increasing interest for gravity theories within spacetimes with large extra dimensions and
the idea that our universe may be a four dimensional singular hypersurface, or ``brane", in a five dimensional spacetime,
or ``bulk". The Randall-Sundrum scenario [1], where our  universe is a four dimensional quasi-Minkowskian edge of a
double-sided perturbed anti-de Sitter spacetime, was the first explicit model where the linearized Einstein equations
were found to hold on the brane, apart from small $1/r^2$ corrections to Newton's potential. This claim was
thoroughly analyzed and the corrections to Newton's law exactly calculated  [2]. Cosmological models were
then built, where the brane is taken to be a Robertson-Walker spacetime embedded in an anti-de Sitter bulk [3]. The
perturbations of these models, in the view of calculating the microwave background
anisotropies, are currently being studied and compared to the perturbations of standard, four dimensional,
Friedmann universes [4-5]. 

In all these papers, which deal with linear perturbations of either a Minkowski or a Robertson-Walker brane in an
anti-de Sitter bulk, corrections to the standard linearized four dimensional Einstein equations are found and analyzed.

On another hand, there exist a number of papers which study under which conditions on the brane and the bulk one
can or cannot recover the exact (and henceforth linearized) four dimensional Einstein equations on the brane. For example
Chamblin Hawking and Reall [6] showed that the exact Schwarzschild solution can hold on a brane. The question
was studied more generally in e.g. [7-8].

Now, if the exact Schwarzschild solution can hold on a brane, a fortiori the linearized Schwarzschild solution and hence
Newton's law can hold as well, and this is in contradiction with the result that there should be  $1/r^2$ corrections to
Newton's potential. A possible explanation for these seemingly contradictory results{\footnote* {Jaume Garriga and
Keichi Maeda, private communications}} is that the  Chamblin-Hawking-Reall
solution exhibits a string-like curvature singularity on a line perpendicular to the brane whereas the Randall-Sundrum
solution is regular everywhere in the bulk. This is true but the question then becomes~: how did Randall-Sundrum find
only one solution (well behaved in the bulk and exhibiting a $1/r^2$ correction to Newton's law in the brane) and did
not find a whole family of solutions, among which theirs plus the linearized version of the Chamblin-Hawking-Reall
solution which is strictly Newtonian on the brane.

The answer obviously lies in an analysis of the boundary conditions far away from the brane which the various
authors choose for the perturbations. Since some authors [1-2] use a field theoretic approach to study the
perturbation equations, based on Green's functions and retarded propagators, whereas some others [5-7] use a geometric
approach which stems from the theory of thin shells in general relativity, it is not completely straightforward to see
where the two approaches differ and where different choices of boundary conditions are made. It is the purpose of this
paper to reconcile the two points of view.

In a first step (Section II) we analyze the example of a massless scalar field in a five dimensional Minkowski
bulk which is simpler than, but akin to, the Randall-Sundrum scenario. We show that the standard field theoretic approach
yields a solution which is well-behaved far away in the bulk but which is non-``newtonian" on the brane. We then show
that the price to pay to get a ``newtonian" solution on the brane is to choose a solution which diverges at infinity in the
bulk.

In Section III we show that the situation is similar in the Randall-Sundrum scenario, with the important
difference that perturbations which diverge at infinity in the bulk must not be necessarily  rejected as their
divergence may be only a gauge effect. Indeed, we show in Section IV that the well-behaved (at least outside the source) 
linearized Chamblin-Hawking-Reall solution corresponds to perturbations which diverge in the conformally Minkowskian
coordinate system used in the field theoretic approach. Finally, in Section V, we rederive the now standard $1/r^2$
correction to Newton's law obtained in [1-2] but with an emphasis on the choice of boundary condition which yields the
result.

We draw a few conclusions in Section VI.

\bigskip
\noindent
{\csc II. The scalar field analogy}
\medskip

Consider a 5-dimensional Minkowski space-time in Minkowskian coordinates $(x^0, \vec r=(x^1, x^2, x^3), y)$~; cut it
along the timelike hypersurface $y=0$~; make a copy of the $y\geq0$ region~; paste it along the original and get a
``$Z_2$-symmetric bulk", that is a double-sided ``half" 5-D Minkowski space-time.

Consider now a massless scalar field $\Phi(x^0,\vec r,y)$ in this bulk, obeying the Klein-Gordon equation
$$\square_5\Phi=0\eqno (2.1)$$
everywhere but on the ``brane" $y=0$, the boundary condition being
$$\partial_y\Phi|_0=\alpha\delta_3(\vec r)\eqno(2.2)$$
with $\alpha$  a constant.

In the analogy we are developping here, the  5-D Minkowski bulk replaces the anti-de Sitter bulk of the Randall-Sundrum
scenario~; Equation (2.1) replaces the linearized Einstein equations governing the perturbations of the anti-de Sitter
bulk~; the hypersurface $y=0$ represents the brane~; and equation (2.2) replaces the Israel junction conditions, in the
case when ``matter" on the brane is a static, point-like source.

Since we consider a static source, we shall look for a static solution of (2.1).

A standard way  to solve (2.1) and (2.2) in the static case is to replace them by the equation, valid  for all $\vec r$ and
all $y$ (positive, zero or negative)
$$\triangle_4\Phi(\vec r,y)=2\alpha\delta_3(\vec r)\delta(y)=2\alpha\delta_4(\vec r,y)\eqno (2.3)$$
which embodies the bulk ``$Z_2$-symmetry". (This is how Garriga and Tanaka [2] for example solve the corresponding
equations of the Randall-Sundrum scenario.) Equation (2.3) is easily solved within the theory of distributions and
yields{\footnote* {One can also solve this equation in Fourier space. One decomposes $\Phi(\vec r,y)$ as $\Phi(\vec
r,y)=\int\! {d^3\vec k\over(2\pi)^{{3\over2}}} e^{i\vec k.\vec r}\hat\Phi_{\vec k}(y)$. Using the fact that the Fourier
transform of $\delta_n(\vec r)$ is $(2\pi)^{-{n\over2}}$ the equation for $\hat\Phi_{\vec k}(y)$ is~: $-k^2\hat\Phi_{\vec
k}+\partial^2_{yy}\hat\Phi_{\vec k}={2\alpha\over(2\pi)^{3\over2}}\delta(y)$. The solution is obtained by
further decomposing $\hat\Phi_{\vec k}(y)$ in a ``tower of massive modes" as
$\hat\Phi_{\vec k}(y)=\int\!{dm\over\sqrt{2\pi}}e^{imy}\bar\Phi_{\vec k,m}$ with
$\bar\Phi_{\vec k,m}=-{2\alpha\over(2\pi)^2}{1\over k^2+m^2}$. Hence the solution, after regularization~:
$\Phi(\vec r,y)=-{\alpha\over2\pi^2}{1\over r^2+y^2}$. }}
$$\Phi(\vec r, y)=-{\alpha\over2\pi^2}{1\over (r^2+y^2)}\,.\eqno(2.4)$$
On the brane, the solution $\Phi (\vec r, y=0)\propto 1/r^2$ is ``non-newtonian" (in the Randall-Sundrum set up the
corresponding solution will incorporate the $1/r^2$ corrections to Newton'a law)~; as for
$\partial_y\Phi|_0$, it is zero everywhere, but at $\vec r=0$ where it is a delta function (see (2.2)). Near the brane, the
expansion of (2.4), in a distributional sense, is
$$\Phi(\vec r, y)=-{\alpha\over2\pi^2}{1\over r^2}+\alpha\delta_3(\vec r)\,y+{\alpha\over2\pi^2}{1\over r^4}\, y^2
-\alpha\triangle\delta_3(\vec r)\,{y^3\over6}+{\cal O}(y^4)\eqno(2.5)$$
and exhibits delta-like singularities at $\vec r=0$ for all $y\neq0$ which are an artifact of the expansion since they
disappear when the series is summed to give back (2.4).

Before trying to see how one could get a ``newtonian" $1/r$ solution on the brane, let us solve the equations (2.1) and
(2.2) in a slightly different way. Everywhere outside the brane, and for $y>0$, equation (2.1) reduces, in the static case,
to
$\triangle_4\Phi(\vec r,y)=0$. Let us solve this equation in Fourier space~: $\Phi(\vec r,y)=\int\! {d^3\vec
k\over(2\pi)^{{3\over2}}} e^{i\vec k.\vec r}\hat\Phi_{\vec k}(y)$ with  
$$-k^2\hat\Phi_{\vec k}+\partial^2_{yy}\hat\Phi_{\vec k}=0\,.\eqno(2.6)$$
(In the Randall-Sundrum scenario this equation will become a Bessel equation.)  This equation has a well behaved
solution at $y\to+\infty$ which is 
$$\hat\Phi_{\vec k}={b_{\vec k}\over(2\pi)^{3\over2}}e^{-ky}\,.\eqno(2.7)$$ 
One now imposes the boundary condition (2.2) which gives~: $\partial_y\hat\Phi_{\vec k}|_0=\alpha/(2\pi)^{3\over2}$,
that is $b_{\vec k}=-{\alpha\over k}$. Hence $\Phi(\vec r,y)=-\alpha\int\!  {d^3\vec k\over(2\pi)^3}{1\over k} e^{(i\vec
k.\vec r-ky)}=-{\alpha\over2\pi^2}{1\over r^2+y^2}$. In thus proceeding we see that the ``standard" solution (2.4) is the
one which corresponds to Fourier modes $\hat\Phi_{\vec k}(y)$ which converge when $y\to+\infty$. 

Let us now see what scalar field we must choose in the bulk in order to recover  a ``newtonian" $1/r$ solution on the
brane. In order to do so we impose
$$\Delta_3\Phi|_0=\alpha\delta_3(\vec r)\qquad\Longrightarrow\qquad\Phi(\vec r,
y=0)=-{\alpha\over4\pi}{1\over r}\,.\eqno(2.8)$$
Then the junction condition (2.2) gives the first $y$ derivative of $\Phi(\vec r, y)$ on the brane~; as for the bulk
equation (2.1) together with (2.8) it gives the second $y$ derivative, so that we get by iteration and for $y\geq0$
$$\Phi(\vec r,y)=\alpha\left[-{1\over4\pi r}+\delta_3(\vec
r)\left(y-{y^2\over2}\right)-\triangle_3\delta_3(\vec r)\left({y^3\over6}-{y^4\over24}\right)+{\cal
O}(y^5)\right]\eqno(2.9)$$ which is equal to the ``newtonian" solution everywhere outside the source $\vec r=0$. (In
the Randall-Sundrum set up this solution will become the expansion in $y$, for $\vec r\neq0$, of the linearized
Chamblin-Hawking-Reall solution.) In Fourier space we have, using the fact that the Fourier transform of $1/4\pi r$ is
$[(2\pi)^{3/2}k^2]^{-1}$
$$\hat\Phi_{\vec k}(y)={\alpha\over(2\pi)^{3\over2}}\left[-{1\over k^2}+\left(y-{y^2\over2}\right)+
k^2\left({y^3\over6}-{y^4\over24}\right)+{\cal
O}(y^5)\right]\,.\eqno(2.10)$$
Now $\hat\Phi_{\vec k}(y)$ is the solution of (2.6) which coincides with the expansion (2.10) near  the brane. One hence
readily obtains $\hat\Phi_{\vec k}(y)$ in closed form as
$$\hat\Phi_{\vec k}(y)={\alpha\over2(2\pi)^{3\over2}}{1\over k^2}\left[(k-1)e^{ky}-(k+1)e^{-ky}\right]
\eqno(2.11)$$
which diverges exponentially as $ky\to+\infty$, so that its Fourier transform $\Phi(\vec r,y)$ (whose expansion near
the brane is given by (2.9)) is not a priori defined.

We therefore see on this scalar field example that the price to pay in order to get a ``newtonian" solution on the brane is
that we must choose a solution whose Fourier transform diverges at infinity in the bulk. The situation will be similar in
the Randall-Sundrum scenario, with the very important difference that perturbations which diverge at infinity in the bulk
will not have to be necessarily  rejected as their divergence may be only a gauge effect.

\bigskip
\noindent
{\csc III. The equations for (static) gravity on a quasi-Minkowskian brane embedded in a perturbed anti-de Sitter bulk}
\medskip

We use conformally Minkowskian coordinates $X^A=\{x^\mu=[x^0, \vec r=(x^1,x^2,x^3)],w\}$ to
describe  a five dimensional perturbed anti-de Sitter spacetime ${\cal V}_5$. We write the metric as
$$ds^2|_5= {\cal G}_{AB}\,dX^AdX^B\quad\hbox{with}\quad {\cal G}_{AB}=\left({{\cal L}\over
w}\right)^2\,(\eta_{AB}+\gamma_{AB})\eqno(3.1)$$ 
 where $\cal L$ is a (positive) constant. 
 ${\cal V}_5$ is taken to be an Einstein space, solution of  the Einstein equations ${\cal R}_{AB}=-{4\over{\cal L}^2}{\cal
G}_{AB}$ where ${\cal R}_{AB}$ is the Ricci tensor of the metric ${\cal G}_{AB}$. In the gauge 
$$ \gamma_{Aw}=0\quad,\quad \gamma^\mu_\mu=\partial_\rho \gamma^\rho_\mu=0\eqno(3.2)$$ 
 these equations reduce to (see e.g. [5] for details)
$$\square_4\gamma_{\mu\nu}+\partial_{ww}\gamma_{\mu\nu}-{3\over
w}\partial_w\gamma_{\mu\nu}=0\,.\eqno(3.3)$$

We consider now in ${\cal V}_5$ the hypersurface $\Sigma$ defined by
$$w={\cal L}+\zeta(x^\mu)\eqno(3.4)$$
where the function $\zeta(x^\mu)$ is a priori arbitrary and describes the so-called ``brane-bending" effect.
The induced metric on $\Sigma$ is 
$$ds^2=(\eta_{\mu\nu}+h_{\mu\nu})dx^\mu dx^\nu\quad\hbox{with}\quad h_{\mu\nu}=\gamma_{\mu\nu}|_{\Sigma} -
2{\zeta\over{\cal L}}\eta_{\mu\nu}\,.\eqno(3.5)$$

The Randall-Sundrum brane scenario is obtained by cutting ${\cal V}_5$ along $\Sigma$, by making a copy of the
$w\geq{\cal L}+\zeta$ side and pasting it along  $\Sigma$. The integration of Einstein's equations across the edge, or
brane,  of this new manifold yields the Lanczos-Darmois-Israel equations (see e.g. [5] for details) which give the
stress-energy tensor of the matter in the brane $\Sigma$ as
$\kappa T^\mu_\nu=-{6\over{\cal L}}\delta^\mu_\nu+\kappa  S^\mu_\nu$
with $\kappa$ a coupling constant and
$${\kappa\over2}\left( S_{\mu\nu}-{1\over3}\eta_{\mu\nu} S\right)=\partial_{\mu\nu}\zeta-{1\over2}
(\partial_w\gamma_{\mu\nu})|_\Sigma\,,\eqno(3.6)$$
which implies
$$\square_4\zeta=-{\kappa\over6}S\,.\eqno(3.7)$$
Equations (3.3) (3.5) (3.6)  and (3.7) are standard~: they can be found in various guise in the literature [2]. 

In the following we shall concentrate on a static, point-like source
$$S_{00}=M\delta(\vec r)\qquad,\qquad S_{0i}=S_{ij}=0\,.\eqno(3.8)$$
Equation (3.7) then gives $\zeta$ as
$$\zeta=-{\kappa M\over24\pi}{1\over r}\,.\eqno(3.9)$$
The junction condistion (3.6) then gives the first $w$-derivative of the bulk metric on the brane as
$$\partial_w\gamma_{00}|_\Sigma=-{2\kappa M\over 3}\delta (\vec r)\quad,\quad
\partial_w\gamma_{0i}|_\Sigma=0\quad,\quad
\partial_w\gamma_{ij}|_\Sigma=-{\kappa M\over 3}\delta (\vec r)\delta_{ij}-{\kappa M\over
12\pi}\partial_{ij}{1\over r}\,.\eqno(3.10)$$
Finally  the general solution of (3.3) in the static case is a
superposition of Fourier modes~: $\gamma_{\mu\nu}(x^\mu,w)=\int\! {d^3\vec
k\over(2\pi)^{{3\over2}}} e^{i \vec k\,.\vec r}\hat\gamma_{\mu\nu}(\vec k, w)$ with
$$\hat\gamma_{\mu\nu}(\vec k, w)=w^2
\left[e_{\mu\nu}^{(1)}(\vec k)H_2^{(1)}(ikw)+e_{\mu\nu}^{(2)}(\vec k)H_2^{(2)}(ikw)\right]
\eqno(3.11)$$
where (because of (3.2)) the polarization tensors $e_{\mu\nu}^{(1,2)}(\vec k)$ are transverse and traceless (and thus
have  a priori five freely specifiable components), and  where
$H_2^{(1,2)}(ikw)$ are the Hankel functions of first and second kind and of order $2$. When $w\to+\infty$,
$H_2^{(2)}(ikw)$ diverges exponentially but we do not eliminate it a priori, in keeping with the conclusions of the
previous Section. The junction condition (3.10) therefore determines only a combination of the polarization tensors, to
wit
$$\eqalign{e_{00}^{(1)}(\vec k)H_1^{(1)}(ik{\cal L})+e_{00}^{(2)}(\vec k)H_1^{(2)}(ik{\cal L})&=-{2\kappa M\over 3}
{1\over(2\pi)^{3\over2}}{1\over ik{\cal L}^2}\cr
e_{0i}^{(1)}(\vec k)H_1^{(1)}(ik{\cal L})+e_{0i}^{(2)}(\vec k)H_1^{(2)}(ik{\cal L})&=0\cr
e_{ij}^{(1)}(\vec k)H_1^{(1)}(ik{\cal L})+e_{ij}^{(2)}(\vec k)H_1^{(2)}(ik{\cal L})&=-{\kappa M\over 3}
{1\over(2\pi)^{3\over2}}{1\over ik{\cal L}^2}\left(\delta_{ij}-{k_ik_j\over4\pi k^2}\right)\,.\cr}\eqno(3.12)$$

To summarize~: when matter on the brane is a static, point-like source, the metric on the brane is (3.5) with $\zeta$
given by (3.9) and $\gamma_{\mu\nu}|_\Sigma$ given by (3.11) (with $w={\cal L}$) and the polarization tensors
restricted by condition (3.12). In order to determine the brane metric completely we must add an extra condition, for
example that Einstein's linearized equations be recovered on the brane (Section IV) or that (3.11) converge (Section V).

\bigskip
\noindent
{\csc IV. Choosing a bulk such that Schwarzschild's linearized solution holds on the brane}
\medskip
In order to compare the equations (3.3) (3.5) and (3.6) which govern linearized gravity on the brane 
 to the standard 4-D linearized Einstein equations, we find it convenient to go to harmonic coordinates in the brane.
Technically this means~:
$x^\mu\to
\tilde x^\mu= x^\mu+\epsilon^\mu$ $\quad\Longrightarrow\quad$ $ h_{\mu\nu}\to\tilde h_{\mu\nu}=
h_{\mu\nu}+\partial_\mu\epsilon_\nu+\partial_\nu\epsilon_\mu$, that is  using (3.5)
$$\tilde h_{\mu\nu}=\gamma_{\mu\nu}|_{\Sigma} -
2{\zeta\over{\cal L}}\eta_{\mu\nu}+\partial_\mu\epsilon_\nu+\partial_\nu\epsilon_\mu\,.\eqno(4.1)$$ 
The harmonicity condition, to wit
$\partial_\mu
\left(\tilde h^\mu_\nu-{1\over2}\delta^\mu_\nu\tilde h\right)=0$  imposes
$$\square_4\epsilon_\mu=-{2\over{\cal L}}\partial_\mu\zeta\,.\eqno(4.2)$$
 We then have, using (3.6) and (3.7)
$$\square_4\tilde h_{\mu\nu}=-{2\over{\cal L}}\kappa\left( S_{\mu\nu}-{1\over2}\eta_{\mu\nu} S\right)
+\square_4\gamma_{\mu\nu}|_\Sigma-{2\over{\cal L}}(\partial_w\gamma_{\mu\nu})|_\Sigma\,.\eqno(4.3)$$
If the sum of the last two terms is zero the linearized Einstein equations are recovered on the brane, it being
understood that ${\cal L}^{-1}\kappa\equiv 8\pi G$, $G$ being Newton's constant.

Let us again concentrate on the case of a static and point-like source (3.8).

The linearized Einstein equations $\square_4\tilde h_{\mu\nu}=-{2\over{\cal L}}\kappa\left(
S_{\mu\nu}-{1\over2}\eta_{\mu\nu} S\right)$ then give
$$\tilde h_{00}={\kappa M\over4\pi{\cal L}}\,{1\over r}={2GM\over
r}\quad,\quad\tilde h_{0i}=0\quad,\quad\tilde h_{ij}={\kappa M\over4\pi{\cal L}}\,{1\over r}\delta_{ij}=
{2GM\over r}\delta_{ij}\eqno(4.4)$$
which is nothing but the expansion at linear order of the Schwarzschild metric in harmonic coordinates.

Let us now see which bulk is required to yield (4.4).
As an easy manipulation of the Bessel functions shows the junction condition (3.12) together with the condition that
Einstein's linearized equations be recovered on the brane, that is 
$$\triangle_3\gamma_{\mu\nu}|_\Sigma-{2\over{\cal L}}(\partial_w\gamma_{\mu\nu})|_\Sigma=0$$
completely determine the polarisation vectors entering $\hat\gamma_{\mu\nu}(\vec k,w)$ in (3.11) and we have
$$\hat\gamma_{\mu\nu}(\vec k,w)=i{\kappa\pi M\over12{\cal L}}{1\over(2\pi)^{3\over2}}\,w^2
\left[H_0^{(1)}(ik{\cal L})H_2^{(2)}(ikw)-H_0^{(2)}(ik{\cal L})H_2^{(1)}(ikw)\right]c_{\mu\nu}\eqno(4.5)$$
with $c_{00}=2$, $c_{0i}=0$ and $c_{ij}=\delta_{ij}-k_ik_j/4\pi k^2$.

These Fourier modes diverge exponentially when $kw\to+\infty$ so that their Fourier integral $\gamma_{\mu\nu}(\vec
r,w)$ is not defined. However we do know the geometry of the bulk when the metric on the brane is given by (4.4). It is
the expansion at linear order of the Chamblin-Hawking-Reall  metric [6] which reads, in Gaussian normal coordinates
$(\tilde x^\mu,y)$, everywhere outside the source, that is for $\tilde x^i\neq0$
$$ds^2|_5=dy^2+e^{-2y/{\cal L}}\left[-\left(1-{2GM\over r}\right)(d\tilde x^0)^2+\left(1+{2GM\over
r}\right)\delta_{ij}d\tilde x^i d\tilde x^j\right]\,.\eqno(4.6)$$ Therefore the badly divergent metric (4.5) must turn into
the good-looking metric (4.6), at least outside the source, when one goes from the conformally Minkowskian
coordinates $(\tilde x^\mu,w)$ to the Gaussian coordinates
$(\tilde x^\mu,y)$. This is indeed what happens, as is shown in the Appendix. 

The conclusion which we can draw from this Section is that divergent solutions of the anti-de Sitter perturbation 
equations (3.3) must perhaps not be a priori be rejected as they can in some cases be transformed into ``regular" ones by
a mere change of coordinates. We write ``regular" with inverted commas because the Chamblin-Hawking-Reall  metric
possesses a curvature singularity at $\vec r=0$, for all $y$, and may have to be rejected on these grounds. But this is a
separate issue from the one we are discussing here which deals with the properties of the bulk far away from the brane.

\bigskip
\noindent
{\csc V. On the $1/r^2$ correction to Newton's law}
\medskip
We just saw that in order to recover Einstein's linearized  equations (and hence Newton's) on the brane one must allow for
divergent perturbations in the bulk, at least when using conformally Minkowskian coordinates. We show in this Section
that the now standard $1/r^2$ correction to Newton's law is due to imposing perturbations in the bulk which remain
small in conformally Minkowskian coordinates. 

The ``standard" way to study linearized gravity on a quasi-Minkowskian brane is to combine the perturbation
equation (3.3) and the junction condition (3.5) into the single equation, valid for all $w$
$$\square_4\gamma_{\mu\nu}+\partial_{ww}\gamma_{\mu\nu}-{3\over
w}\partial_w\gamma_{\mu\nu}=-2\kappa\delta(w-{\cal
L})\left(S_{\mu\nu}-{1\over3}\eta_{\mu\nu}S-{2\over\kappa}\partial_{\mu\nu}\zeta\right)\eqno(5.1)$$
and to solve it by means of retarded propagators. As is now well known this calculation leads to small corrections to
Newton's law of gravity, see [1-2].

We shall recover here this result in a slightly different way which shows that it stems from a particular
choice of boundary condition far away from the brane.

As we saw in Section I, solving (5.1) in a distributional sense amounts, in the static case at least,  to choosing the
solution of (3.3) which converges far from the brane, that is to choosing $e_{\mu\nu}^{(2)}(\vec k)=0$ in (3.11). With this
choice of boundary condition at $w\to+\infty$, the junction conditions (3.12) completely determine the remaining
polarization tensor $e_{\mu\nu}^{(1)}(\vec k)$. The bulk metric  is then known as~:
$$\gamma_{\mu\nu}(\vec r,w)=\int\!{d^3k\over(2\pi)^{3\over2}}\,e^{i\vec k\,.\vec r}\hat\gamma_{\mu\nu}(\vec k,
w)\quad ,\quad\hat\gamma_{\mu\nu}(\vec k,w)=
{\kappa M\over3{\cal L}(2\pi)^{3\over2}}
\,w^2\,{K_2(kw)\over k{\cal L}\,K_1(k{\cal L})}\,c_{\mu\nu}\eqno(5.2)$$
with $c_{00}=2$, $c_{0i}=0$ and $c_{ij}=\delta_{ij}-k_ik_j/4\pi k^2$ and where $K_\nu(z)$ is the modified Bessel
function defined as $K_\nu(z)=i{\pi\over2}e^{i\nu{\pi\over2}}H_\nu^{(1)}(iz)$. Near the brane this metric reduces
to, setting
$u={w\over{\cal L}}-1$
$$\hat\gamma_{\mu\nu}(\vec k,w)=
{\kappa M{\cal L}\over3(2\pi)^{3\over2}}
\left\{{K_2(k{\cal L})\over k{\cal L}\,K_1(k{\cal L})}-u-{k{\cal L}\over4}u^2\left[{K_2(k{\cal L})\over K_1(k{\cal
L})}-3{K_0(k{\cal L})\over K_1(k{\cal L})}\right]+{\cal O}(u^3)\right\}c_{\mu\nu}\eqno(5.3)$$
(which implies in particular $\partial_w\gamma_{00}|_\Sigma=-{2\kappa M\over 3}\delta_3(\vec r)$ in accordance
with (3.10). The appearance of Dirac distributions in the expansion of $\gamma_{\mu\nu}(\vec r,w)$ does not however
necessarily mean that $\gamma_{\mu\nu}(\vec r,w)$ is singular at $\vec r=0$ as the sum may be regular, as in the
scalar field example treated in Section 2).

Let us now concentrate on the $h_{00}$ component of the metric on the brane (which is the same in the $x^\mu$
coordinates and the harmonic coordinates $\tilde x^\mu$). With $\zeta$ given by (3.9), and hence $\hat\zeta=-{\kappa
M\over6k^2}(2\pi)^{-{3\over2}}$, it reads
$$\hat h_{00}(\vec k)=\hat{\tilde h}_{00}(\vec k)=\hat\gamma_{00}|_\Sigma+2{\hat\zeta\over{\cal L}}=
{\kappa M\over k^2{\cal L}(2\pi)^{3\over2}}\left[1-{2k{\cal L}\over3}{K_0(k{\cal L})\over K_1(k{\cal L})}\right]\,.
\eqno (5.4)$$
Taking the Fourier transform and integrating over angles we obtain, setting $\alpha=r/{\cal L}$
$$h_{00}(\vec r)={\kappa M\over4\pi{\cal L}}{1\over r}\left(1+{4\pi\over3}{\cal K}_\alpha\right)\quad\hbox{with}\quad
{\cal K}_\alpha=\lim_{\epsilon\to0}\int_0^{+\infty}\! du\,\sin(u\alpha)\,{K_0(u)\over K_1(u)}e^{-\epsilon
u}\,.\eqno(5.5)$$ It is a (fairly) straightforward exercise to see that $\lim_{\alpha\to0}{\cal
K}_\alpha=\alpha^{-1}={\cal L}/r$ and that
$\lim_{\alpha\to\infty}{\cal K}_\alpha=\pi/2\alpha^2=\pi ({\cal L}/r)^2/2$. We hence recover that at short distances
the correction to Newton's law is in ${\cal L}/r$, whereas as distances large compared with the characteristic scale
${\cal L}$ of the anti-de Sitter bulk the correction is reduced by another ${\cal L}/r$ factor, in agreement with [1-2]
$$\lim_{r/{\cal L}\to\infty}h_{00}(\vec r)={2GM\over r}\left[1+{2\over3}\left({{\cal L}\over
r}\right)^2\right]\,.\eqno(5.6)$$

\bigskip
\noindent
{\csc VI. Conclusions}
\medskip
We showed in this paper that the commonly accepted $1/r^2$ correction to Newton's law of the Randall-Sundrum
scenario arises from the fact that one imposes the bulk perturbations to be bounded in a conformally
Minkowskian coordinate system. In a Gaussian normal coordinate system, which differs more and more from a 
conformally Minkowskian one as one goes further and further away from the brane, one can have bounded perturbations
together with a strictly Newtonian potential on the brane (this is the linearized Chamblin-Hawking-Reall solution). 

A question therefore arises~: how can one disantangle gauge effects in order to allow only for {\it geometrically}
bounded perturbations in the anti-de Sitter bulk~? We think that this question can be answered by a careful analysis of
the asymptotics of anti-de Sitter spacetime, using Schwarzschild-like coordinates adapted to its universal covering.

Another question one may ask is~: when considering an extended (static source) to the Chamblin-Hawking-Reall metric,
what is the solution like inside the source~? Do we obtain a (unstable)  curvature singularity when its size tends to zero
as in the Chamblin-Hawking-Reall solution which allows us  to reject this solution on these grounds~? This is
not completely clear as, first, a linear analysis is no longer sufficient and, second, it is known that when the energy
density becomes high enough in the brane Einstein equations cannot hold  (see e.g. [8]). We leave this problem to another
work [9].

\bigskip
\noindent
{\csc Acknowledgements}
\bigskip
We gratefully acknowledge very fruitful discussions with Jaume Garriga, Gilles Espo\-sito-Farese, Keichi Maeda and
Takahiro Tanaka.

\bigskip
\noindent
{\csc Appendix}
\medskip
We show here the equivalence between the metric (3.1) (3.2) (4.5)  and the metric (4.6) outside the source.

Consider the perturbed anti-de Sitter metric in conformally Minkowskian coordinates $X^A=\{x^\mu=[x^0,\vec
r=(x^i)],w\}$
$$ds^2|_5=\left({{\cal L}\over w}\right)^2(\eta_{AB}+\gamma_{AB})\eqno(A.1)$$
with $\gamma_{Aw}=\gamma^\mu_\mu=\partial_\rho\gamma^\rho_\mu=0$, with $\gamma_{\mu\nu}(x^\mu,w)
=\int{d^3k\over(2\pi)^{3\over2}}e^{i\vec k\,.\vec r}\hat\gamma_{\mu\nu}(\vec k, w)$ and
$$\hat\gamma_{\mu\nu}(\vec k,w)=i{\kappa\pi M\over12{\cal L}}{1\over(2\pi)^{3\over2}}\,w^2
\left[H_0^{(1)}(ik{\cal L})H_2^{(2)}(ikw)-H_0^{(2)}(ik{\cal L})H_2^{(1)}(ikw)\right]c_{\mu\nu}\eqno(A.2)$$
where $H_\nu^{(a)}(z)$ are Hankel functions of order $\nu$, of kind $(a)$ and argument $z$, and where
$c_{00}=2$, $c_{0i}=0$ and $c_{ij}=\delta_{ij}-k_ik_j/4\pi k^2$. As an easy manipulation of Bessel functions shows, 
this metric reduces near $w={\cal L}$ to 
$$\eqalign{\gamma_{00}(x^\mu,w)&={\kappa M\over3\pi{\cal L}}{1\over r}+{\kappa M{\cal L}\over3}\delta_3(\vec
r)\left[1-\left({w\over{\cal L}}\right)^2\right]+...\cr
\gamma_{0i}(x^\mu,w)&=0\cr
\gamma_{ij}(x^\mu,w)&={\kappa M\over12\pi{\cal L}}{1\over r}\left(\delta_{ij}+{x_ix_j\over r^2}\right)
+{\kappa M{\cal L}\over6}\left(\delta_{ij}\delta_3(\vec r)+{1\over4\pi}\partial_{ij}{1\over
r}\right)\left[1-\left({w\over{\cal L}}\right)^2\right] +...\cr}\eqno(A.3)$$

 The brane is defined by $w={\cal L}+\zeta$ with
$\zeta=-\kappa M/24\pi r$ and the metric on the brane is $ds^2=(\eta_{\mu\nu}+h_{\mu\nu})dx^\mu dx^\nu$ with 
$h_{\mu\nu}=\gamma_{\mu\nu}|_{w={\cal L}}-2\zeta\eta_{\mu\nu}/{\cal L}$. As shown in the main text this metric on
the brane is nothing but the linearized Schwarszchild solution which can be written in the familiar form  $\tilde
h_{00}=\kappa M/4\pi{\cal L}r$, $\tilde h_{0i}=0$, $\tilde h_{ij}= \delta_{ij}\kappa M/4\pi{\cal L}r$ when going to
harmonic coordinates
$x^\mu\to\tilde x^\mu$ on the brane (N.B.~: in both coordinate systems the $(00)$ components are the same,
$h_{00}=\tilde h_{00}$).

Let us now perform an infinitesimal change of coordinates $X^A\to \bar X^A=X^A+\bar\epsilon^A$. If we choose
$$\bar\epsilon^4=-{w\zeta\over{\cal L}}\qquad,\qquad \bar\epsilon_\mu=-{{\cal
L}\over2}\partial_\mu\zeta\left[1-\left(w\over{\cal L}\right)^2\right]\eqno(A.4)$$
then the new coordinates will be Gaussian normal, that is the brane will be located at $\bar w={\cal L}$ and the metric
will read
$ds^2|_5=\left({{\cal L}\over\bar w}\right)^2(\eta_{AB}+\bar\gamma_{AB})d\bar X^Ad\bar X^B$
with
$$\eqalign{\bar \gamma_{00}&={\kappa M\over4\pi{\cal L}}{1\over r}+{\kappa M{\cal L}\over3}\delta_3(\vec
r)\left[1-\left(w\over{\cal L}\right)^2\right]+...\cr
\bar \gamma_{0i}&=0\cr
\bar \gamma_{ij}&={\kappa M\over12\pi{\cal L}}{1\over r}\left(2\delta_{ij}+{x_ix_j\over r^2}\right)
+{\kappa M{\cal L}\over6}\delta_3(\vec r)\delta_{ij}\left[1-\left(w\over{\cal L}\right)^2\right]+...\cr}\eqno(A.5)$$

We now set ${\bar w\over{\cal L}}=e^{y\over{\cal L}}$, and we also go to harmonic coordinates $\tilde x^\mu$ on the
brane, that is we perform the  infinitesimal change of coordinates $\bar x^\mu=\tilde x^\mu-\tilde\epsilon^\mu$ with
$\tilde\epsilon_0=0$ and $\tilde\epsilon_i={\kappa M\over24\pi{\cal L}}{x_i\over r}$ so that we have
$\partial_\mu(\tilde h^\mu_\nu-\delta^\mu_\nu\tilde h)=0$. The bulk metric then reads
$$ds^2|_5=dy^2+e^{-2{y\over{\cal L}}}(\eta_{\mu\nu}+\tilde\gamma_{\mu\nu})d\tilde x^\mu d\tilde x^\nu\eqno(A.6)$$
with
$$\eqalign{\tilde\gamma_{00}&={\kappa M\over4\pi{\cal L}}{1\over r}-{2\kappa M\over3}\delta_3(\vec
r)y\left(1+{y\over{\cal L}}\right)+{\cal O}(y^3)\cr
\tilde\gamma_{0i}&=0\cr
\tilde\gamma_{ij}&={\kappa M\over4\pi{\cal L}}{1\over r}\delta_{ij}-{\kappa M\over3}\delta_3(\vec
r)\,y\left(1+{y\over{\cal L}}\right)\delta_{ij}+{\cal O}(y^3)\cr}\,.\eqno(A.7)$$
Setting $\kappa/{\cal L}=8\pi G$ we therefore see that, everywhere outside the source, that is for $\vec r\neq0$ the
metric (A.7) is nothing but the linearized Chamblin-Hawking-Reall solution.

We showed the equivalence of the two metrics near the brane. But an iteration of the bulk equations (3.3) would allow us
to extend the transformation to the whole bulk (see [8] for an example of such a procedure). A important point to note is
that, see (A.4), the transformation cannot be infinitesimal for large $w$. This is not surprising as conformally
Minkowskian and Gaussian normal coordinates differ more and more as we go away from the brane. This explains also
why the two metrics look so different and why one is even divergent when the other is perfectly well behaved, at least
outside the source.

\bigskip\bigskip
\noindent
{\csc References}
\bigskip
\item{[1]} L. Randall, R. Sundrum, Phys. Rev. Lett. 83 (1999) , 4690

\item{[2]} J. Garriga, T. Tanaka,  Phys. Rev. Letters 84 (2000) 2778 ;
 C. Csaki, J. Erlich, T.J. Hollowood, Y. Shirman, Nucl. Phys. B581 (2000) 309; S.B. Giddings, E. Katz, L. Randall, JHEP
0003 (2000) 023; C. Csaki, J. Erlich, T.J. Hollowood, Phys. Rev. Lett. 84 (2000) 5932; C. Csaki, J. Erlich, T.J. Hollowood,
Phys. Lett. B481 (2000) 107; I.Ya. Aref'eva, M.G. Ivanov, W. Muck, K.S. Viswanathan, I.V. Volovich, Nucl. Phys. B590 (2000)
273;  Z. Kakushadze, Phys. Lett. B497 (2000) 125

\item{[3]} P. Bin\'etruy, C. Deffayet, U. Ellwanger,  D. Langlois,  Phys. Lett. B477, 285 (2000);
 P. Kraus, JHEP 9912 (1999) 011, S. Mukoyama,  Phys. Lett. B473 (2000) 241;
 D.N. Vollick, C.Q.G. 18 (2001) 1; D. Ida, JHEP 0009 (2000) 014;  S. Mukohyama, T. Shiromizu, K. Maeda, Phys. Rev.
D62 (2000) 024028;  R. Maartens, Phys. Rev. D62 (2000) 084023; P. Kanti, K.A. Olive, M. Pospelov, Phys. Lett. B468
(1999) 31

\item{[4]} S. Mukohyama, Phys. Rev. D62 (2000) 084015; H. Kodama, A. Ishibashi and O. Seto, Phys. Rev. D62
(2000) 064022;  D. Langlois, Phys. Rev. D62 (2000)126012;  C. van de Bruck, M. Dorca, R. Brandenberger, A. Lukas,
Phys. Rev. D62 (2000) 123515; K. Koyama, J. Soda, Phys. Rev. D62 (2000) 123502;  D. Langlois, R. Maartens, D. Wands,
Phys. Lett. B489 (2000) 259;  S. Mukohyama, CQG 17 (2000) 4777; C. Gordon, D. Wands, B. Bassett, R. Maartens, 
Phys. Rev. D63 (20001) 123506;  C. van de Bruck, M. Dorca, ``On cosmological perturbations on a brane in an anti-de
Sitter bulk", hep-th/0012073; M. Dorca, C. van de Bruck. ``Cosmological perturbations in brane worlds~: brane bending and
anisotropic stresses", hep-th/0012116

\item{[5]} N. Deruelle, T. Dolezel, J. Katz, ``Perturbations of brane
world", hep-th/0010215, Phys. Rev. D63 (2001) 083513 

\item{[6]} Chamblin, Hawking and Reall, Phys. Rev. D61 (2000) 065007

\item{[7]} T. Shiromizu, K. Maeda, M. Sasaki, Phys. Rev. D62 (2000) 024012;  M. Sasaki, T, Shiromizu, K. Maeda,  Phys. Rev.
D62 (2000) 024008; R. Maartens, ``Geometry and dynamics of brane worlds", gr-qc/0101059

\item{[8]}  N. Deruelle, Joseph Katz, ``Gravity on branes",

\item{[9]} N. Deruelle, Joseph Katz, in preparation.
\end